\documentclass[11pt]{amsart} %
\usepackage{amsfonts} 
\usepackage{epsfig}
\usepackage{hhline,eufrak,euscript,delarray}

\newtheorem{theorem}{Theorem}[section]
\newtheorem{lemma}[theorem]{Lemma}

\author{\textsc{Vlady} RAVELOMANANA }
\author{\textsc{Alphonse Laza} RIJAMAMY}
\email{ vlad@lipn.univ-paris13.fr, rilazako@yahoo.fr}
\address{Vlady RAVELOMANANA  (corresponding author), LIPN, Universit\'e de Paris-Nord,
 F 93430 Villetaneuse, France.}
\address{Alphonse Laza RIJAMAMY, 
Universit\'e de Madagascar, Ankatso -- Tana 101, Madagascar.}

\title{Creation and Growth of Components in a Random Hypergraph Process}

\def\qE{\mathbb{E}}
\def\a{\xi}

\def\Vz{{\vartheta}_z}

\def \beq{\begin{equation}}
\def \eeq{\end{equation}}
\def \be{\begin{eqnarray*}}
\def \ee{\end{eqnarray*}}
\def \ben{\begin{eqnarray}}
\def \een{\end{eqnarray}}

\def \b{\big}

\def\coeff#1{\left[ #1 \right]}
\def\point{\noindent $\, \, \, \mathbf{\bullet}\, \, \,$}

\newcommand{\ENDPROOF}{\hfill \qed}

\def\Ht{${\{\mathbb{H}(n,t)\}}_{0 \leq t\leq 1} \,$}
\def\VMAX{{V_{\ell}}^{max}}
\def\VL{{V_{\ell}}}

\def \ten{\rightarrow}
\def\qE{\mathbb{E}}

\begin{document}

\begin{abstract}
Denote by an $\ell$-component a connected $b$-uniform hypergraph with 
$k$ edges and $k(b-1) - \ell$ vertices. 
We prove that the expected number of creations of $\ell$-component
during a random hypergraph process tends to $1$ as $\ell$ and $b$ tend
to $\infty$ with the total number of vertices $n$ such that 
$\ell = o\left( \sqrt[3]{\frac{n}{b}} \right)$.
Under the same conditions, we also show 
that the expected number of vertices that ever belong to an
$\ell$-component is approximately $12^{1/3} (b-1)^{1/3} \ell^{1/3} n^{2/3}$.
As an immediate consequence, it follows
that with high probability the largest $\ell$-component during the process is
of size $O( (b-1)^{1/3} \ell^{1/3} n^{2/3} )$. Our results give insight about
the size of giant components inside the phase transition of random hypergraphs.
\end{abstract}


\maketitle

\section{Introduction}
A \textit{hypergraph} $\mathcal{H}$ is a pair $(\mathcal{V}, \mathcal{E})$ where 
$\mathcal{V}=\{1,\, 2, \, \cdots ,\, n\}$ denotes the set of 
vertices of $\mathcal{H}$ and $\mathcal{E}$ is
a family of subsets of $\mathcal{V}$ called edges (or hyperedges). For a general
treatise on hypergraphs, we refer to Berge \cite{BERGE}. We say that
$\mathcal{H}$ is \textit{$b$-uniform} (or simply \textit{uniform}) if for every edge 
$e \in \mathcal{E}$, $|e| = b$. In this paper, all considered hypergraphs
are $b$-uniform. We will study the growth of size and complexity of
connected components of a random hypergraph process \Ht defined as follows.
Let $K_n$ be the complete hypergraph built with $n$ vertices and 
${n \choose b}$ edges (self-loops and multiple edges
are not allowed). \Ht may be constructed by letting each edge $e$ of 
$K_n$ (amongst the ${n \choose b}$ possible edges)
 appear at random time $T_e$, with $T_e$ independent and uniformly
distributed on $(0,\, 1)$ and letting \Ht contain the edges such that
$T_e \leq t$ (for the random graph counterpart of this model, 
we refer the reader to \cite{Ja2000,Ra06}). This model is closely related to
 $\{\mathbb{H}(n, \, M)\}$ where $M \in \coeff{1, \, {n \choose b}}$ 
represents the number of edges picked uniformly at random
amongst the ${n \choose b}$ possible edges and which are
 present in the random hypergraph. The main difference between 
 $\{\mathbb{H}(n,M)\}_{0\leq M \leq {n \choose b}}$ and \Ht is that 
in   $\{\mathbb{H}(n,M)\}_{0\leq M \leq {n \choose b}}$, edges are added at fixed (slotted)
times $1, 2, \ldots$, ${n \choose b}$ so at any time $M$ we obtain 
a random graph with $n$ vertices and $M$ edges,
whereas in \Ht the edges are added at random times.
 At time $t=0$, we have a hypergraph with $n$ vertices and $0$ edge,
and as the time advances all edges $e$ with r.v.~ $T_e$
such that  $T_e \leq t$ (where $t$ is the current time), 
 are added to the hypergraph until $t$ reaches $1$
in which case, one obtains the complete hypergraph $K_n$.

\smallskip

We define the {\it excess} (or the {\it complexity}) of 
a connected $b$-uniform hypergraph as (see also \cite{KL97}):
\beq
\mathrm {excess}(\mathcal{H})=  \sum _{ e \in \mathcal{E}} \left(\vert e \vert -1 \right)%
-\vert \mathcal{V} \vert \, =  \vert \mathcal{E} \vert \times (b-1)- \vert \mathcal{V} \vert \, .
\label{DEF_GENERAL_EXCESS}
\eeq
Namely, the complexity (or excess) of connected
components  ranges from $-1$ (hypertrees) to ${n \choose b}(b-1) - n$ (complete hypergraph).
A connected component with excess $\ell$ ($\ell \geq -1$) is called
 an \textit{$\ell$-component}. 
The notion of excess was first used in \cite{Wr77} where the author
 obtained substantial enumerative results in the study of connected graphs
according to the two parameters  number of vertices and number of edges. 
It was also used in enumerative combinatorics and as well as in various
study of random hypergraphs processes\cite{KL97,KL02,AR05}. 

\smallskip

Numerous results have been obtained for random graphs as witnessed by the books
\cite{Bollobas,JLR2000} and the references therein. In comparison, there are
 very few works about random hypergraphs. One of the most significant results 
was obtained by Schmidt-Pruznan and Shamir \cite{SpS85} who studied the component
structure for random hypergraphs. In particular, they proved that if $b \geq 2$, 
$M =cn$ with $c<1/b(b-1)$ then asymptotically almost surely (a.a.s. for short)
 the largest component of $\mathbb{H}(n,M)$ is of order $\log{n}$ and for
$c = 1/b(b-1)$ it has $\theta(n^{2/3})$ vertices and as $c>1/b(b-1)$ a.a.s. 
$\mathbb{H}(n,M)$ has a unique geant component with $\theta(n)$ vertices. This result
generalizes the seminal papers of Erd\"{o}s and R\'enyi who discovered  the
abrupt change in the structure of the random graph $\mathbb{G}(n,M)$ when
$M = cn$ with $c \sim 1/2$. In \cite{KL02},  Karo\'nski and {\L}uczak
 proved limit theorems for the distribution of the size of the largest component of
$\mathbb{H}(n,M)$ at the phase transition, i.e., $M=n/b(b-1) + O(n^{2/3})$.

\smallskip

In this paper, we consider the \textit{continuous time} random hypergraph
process described above and will study the creation and growth of components of excess $\ell$ 
(or $\ell$-components). A connected component which is not a hypertree
is said \textit{multicyclic} (following the terms used by 
our predecessors in \cite{Ja93,Ja2000,JKLP93}). 

\subsection{Definitions.}
We can observe that there are two manners to create a 
new $(\ell+1)$ component during the \Ht process~:\\
\point either by adding an edge between an existing $p$-component (with $p \leq \ell$)
and $(b-q)$ hypertrees (with $0 \leq q \leq b$)
 such that the edge encloses $q$ distinct vertices in the $p$-component, \\
\point or by joining with the last added edge many connected components such that the
number of multicyclic components diminishes.

\smallskip

Observe that in the first case,
 to create an $\ell$-component, we must have $p+q-1 = \ell$. In this case, it is also
important to note that the number of
multicyclic components remains the same after the addition of the last edge. 

\smallskip

The first transition described above will be denoted
$p \ten \ell$ and the second $\oplus_{i} p_i \ten \ell$.
We say that an $\ell$-component is \textit{created} by a transition $p \ten \ell$ 
with $p < \ell$ or by a transition $\oplus_{i} p_i \ten \ell$. 
For $\ell \geq 0$, we say that an $\ell$-component \textit{grows} when it swallows
some hypertrees (transition $\ell \ten \ell$).

\smallskip

\noindent
Following Janson in \cite{Ja2000}, we have two points of view~:\\
\point \textit{The static view.} Let $\mathcal{C}_{\ell}(m)$ denote
the collection of all $\ell$-components in \Ht. Consider
the family $\mathcal{C}_{\ell}^{\star} = \bigcup_m \mathcal{C}_{\ell}(m)$ for
every $\ell$-component that appears at some stage of the continuous process,
ignoring when it appears~: the elements of  $\mathcal{C}_{\ell}^{\star}$ are
called \textit{static $\ell$-components}.\\
\point \textit{The dynamic view.} A connected component can be viewed
as ``the same'' according to its excess even after it has grown by swallowing
some hypertrees (transition $\ell \ten \ell$). Such component whose excess remains
the same can be viewed as a \textit{dynamic $\ell$-component}
 as its size evolves.

\smallskip

\noindent
We define $\VL = \vert \mathcal{V}_{\ell} \vert$ as the number of vertices that at some stage
of the process belong to an $\ell$-component and 
$\VMAX = \max\{\vert V(C) \vert: C \in \mathcal{C}_{\ell}^{\star} \}$ to be the size of
the largest $\ell$-component that ever appears. 

\smallskip

\noindent
Let $\alpha(\ell; k)$ be the expected number of times a new edge is added by means
of the first type of transition $p \ten \ell$ in order
to create an $\ell$-component with $k$ edges (or  with $k \times (b-1) - \ell$
vertices). Note again that in this case, the number of multicyclic components 
of the \Ht process remains the same after the addition of this edge.

\smallskip

\noindent 
Similarly, let $\beta(\ell; k)$ be the expected number of times an edge
is added joining at least two multicyclic components in order to form a newly
$\ell$-component with a total of $k$ edges. In other terms,
$\beta(\ell; k)$ is the expected number of times at least two
multicyclic components and some hypertrees merge to form an $\ell$-component.

\subsection{Our results and outline of the paper.} 
We combine analytic combinatorics \cite{FS+} and
probabilistic theory \cite{JLR2000} to study the extremal characteristics 
of the components of a random hypergraph process inside its phase transition
\cite{KL02} and find that the size of the largest component with
$k$ (hyper)edges and $k(b-1)-\ell$ vertices is of order 
$O( (b-1)^{1/3} \ell^{1/3} n^{2/3} )$. 

\smallskip

This extended abstract is organized as follows. In the next section,
we introduce the general expression of the expectations of several random
variables of our interest. In section 3, the computations of 
the expectations are developped focusing on the particular and
instructive case of unicyclic components.
The last paragraph provides several technical lemmas useful in order to
study the extremal case, i.e. whenever the excess $\ell$ of the component
 is large. We give there methods on how to investigate 
the number of creations of $\ell$-components
as well as the expectation of their size.

\section{Connected components and expectation of transitions}
\subsection{Expected number of transitions}
In this paragraph, we give a general formal expression of the 
expectation of the number of the first (resp. second) type of transitions $\alpha(\ell;k)$
(resp. $\beta(\ell;k)$). 

\smallskip

We have the following lemma which computes the expected number of
transitions $\alpha(\ell;k)$~:

\begin{lemma}
Let $a=k(b-1) - \ell$. Denote by $\rho(a,k)$ the number of
manners to label an $\ell$-component with $a$ vertices 
such that one edge
 -- whose deletion will  not increase the number of
multicyclic components but will suppress the newly created $\ell$-component -- 
is distinguished among the others. Then,
\beq \label{EQ:FORMAL_ALPHA}
\alpha(\ell;k) = {n \choose a} \rho(a,k) \int_{0}^{1} %
t^{k-1} (1-t)^{ {n \choose b} - {n-a \choose b}  - k} \, dt \, . 
\eeq
\end{lemma} 

\begin{proof}
There are ${n \choose a}$ choices of the $a=k(b-1) - \ell$ vertices of the 
newly created $\ell$-component.
By the definition of $\rho(a;k)$, 
there are ${n \choose a} \times \rho(a;k)$ possible $\ell$-components.
The probability that the previous component (the one before obtaining the current
$\ell$-component) belongs to \Ht is given by 
\beq
t^{k-1} (1-t)^{ \sum_{i=1}^{b-1} {n-a \choose i}{a \choose b-i} + {a \choose b} - k + 1} \, 
\eeq
where the summation in the exponent represents the number of edges not present between
the considered component and the rest of the hypergraph.
The conditional probability that the last edge is added during
the time interval $(t, \, t+dt)$ and not earlier is $dt/(1-t)$. Using the identity
\beq
 \sum_{i=1}^{b-1} {n-a \choose i}{a \choose b-i} = {n \choose b} - {n-a \choose b} 
- {a \choose b}
\eeq
and integrating over all times after some algebra, we obtain (\ref{EQ:FORMAL_ALPHA}).
\end{proof}

Similarly, if we let $\tau(a;k)$ to be the number of ways to label an
$\ell$-component with $a=(k-1)-\ell$ vertices and $k$ edges 
such that one edge -- whose suppression augments the 
number of multicyclic connected components -- is distinguished among the others.
Then, $\beta(\ell;k)$ can be computed as for $\alpha(\ell;k)$ using
exactly $\tau(a;k)$ instead of $\rho(a;k)$.

\smallskip

Next, the following lemma gives some asymptotic values
needed when using formula (\ref{EQ:FORMAL_ALPHA}).

\begin{lemma} \label{LEMME_BINOMIAL_INTEGRAL}
Let $a=(b-1)k - \ell$. We have
\ben \label{EQ:BINOMIAL_INTEGRAL}
& & {n \choose a} \int_{0}^{1} %
t^{k-1} (1-t)^{ {n \choose b} - {n-a \choose b}  - k} \, dt
  =  %
\frac{1}{\sqrt{(b-1)} \, n^{\ell}}  %
\frac{k^{(k-1)} \, \coeff{\left(b-1\right)!}^{k}}{\Big(k(b-1)-\ell \Big)^{kb-\ell}} \cr
& \times & \exp{\left(k(b-2)-\ell   - \frac{(b-1)^4 \, k^3}{24 \, n^2} \right) } 
 \times  \left(1+O\left( \frac{b k}{n} + \frac{b^4 k^2}{n^2}+ 
\frac{b^4 k}{n^2}   + \frac{b^4 k^4}{n^3}  +  %
\frac{k}{n^{b-1}b} + \frac{1}{k} \right) \right) \, . \cr
& &  \, 
\een 
\end{lemma}

\begin{proof}
First, using Stirling formula for factorial we get
\beq \label{EQ:STIRLING}
{n \choose a} = \frac{1}{\sqrt{2 \pi a}} \, \frac{n^a\, e^{a}}{a^a} \, %
\exp{\left( - \frac{a^2}{2 n} - \frac{a^3}{6 n^2} + %
O\left(\frac{a^4}{n^3} + \frac{1}{a}\right)\right)}\, .
\eeq

For $(x, \, y) \in \mathbb{N}^2$, we have 
\beq
\int_{0}^{1} t^x (1-t)^y dt = \frac{x!}{y!}{(x+y+1)!} = \frac{1}{(x+y+1) {x+y \choose x}} \,. 
\eeq
Setting $N= {n \choose b} - {n-a \choose b}$, using standard calculus we then obtain
\beq \label{EQ:BIG_N}
N = \frac{n^{(b-1)} a}{(b-1)!} \, %
\left(1 - \frac{a(b-1)}{2n} + \frac{a^2 (b-1)(b-2)}{6n^2} + %
O\left(\frac{b}{n} \right) + O\left(\frac{ab^3}{n^2} + \frac{b^4}{n^2}\right) \right)\, .
\eeq
Now, using the above formulas we easily find that the integral equals
\ben
\frac{1}{N \, {N \choose k-1}} & = & %
\frac{\sqrt{2 \pi k}}{N^k} \, \frac{(k-1)^{(k-1)}}{e^{k-1}} %
\left( 1+ O\left( \frac{k^2}{N} + \frac{1}{k} \right) \right) \cr
       &=& %
\sqrt{\frac{2 \pi}{k}} \, \frac{k^k}{N^k e^k} %
\left( 1+ O\left( \frac{k^2}{N} + \frac{1}{k} \right) \right) \, \cr
 & = & %
\sqrt{\frac{2 \pi}{k}} \, \frac{k^k}{e^k} \, %
\frac{\coeff{\left(b-1\right)!}^{k}}{n^{k(b-1)}{a^{k}}} \, %
\left( 1+ O\left( \frac{k}{n^{b-1} b } + \frac{1}{k} \right) \right) \, \cr %
&\times& \exp{\left(-k \log{ \left(1 - \frac{a(b-1)}{2n} + \frac{a^2 (b-1)(b-2)}{6n^2} + %
O\left(\frac{b}{n} \right) + O\left(\frac{ab^3}{n^2} + \frac{b^4}{n^2}%
 \right)  \right)   }     \right)} \, . \cr
& & \, 
\een
Therefore by replacing $a$ with $k(b-1) - \ell$ and using (\ref{EQ:STIRLING}), it yields
(\ref{EQ:BIG_N})
\be
\frac{{n \choose a}}{N \, {N \choose k-1}} & \sim & %
\frac{1}{\sqrt{(b-1)} \, n^{\ell}}  %
\frac{k^{(k-1)} \, \coeff{\left(b-1\right)!}^{k}}{\Big(k(b-1)-\ell \Big)^{kb-\ell}} %
\exp{\Big(k(b-2)-\ell\Big)}
\exp{\left(   - \frac{(b-1)^4 \, k^3}{24 \, n^2} \right) } \, , %
\ee
where $\sim$ means that the asymptotic equation holds up to a
factor of
\[
1+O\left( \frac{b k}{n} + \frac{b^4 k^2}{n^2}+ 
\frac{b^4 k}{n^2}   + \frac{b^4 k^4}{n^3}  +  \frac{k}{n^{b-1}b} + \frac{1}{k} \right) \, .
\]
\end{proof}

Lemma \ref{LEMME_BINOMIAL_INTEGRAL} tells us that the 
expectations the random variables of interest rely on the asymptotic number
of the considered connected components. 

\subsection{Asymptotic enumeration of connected hypergraphs}
As far as we know there are not so many results about the asymptotic enumeration of
connected uniform hypergraphs. In this paragraph, we recall some of the 
results established independently in \cite{KL97,Coja,AR05} (the three papers
actually use three different methods). In \cite{AR05}, the authors use
the generating functions approach \cite{GF_OLOGY,JKLP93,FS+,Wr77,Wr78,Wr80} 
to count exactly and asymptotically 
 connected labeled  $b$-uniform hypergraphs. If $A(z) = \sum_n a_n z^n$ 
and $B(z) = \sum_n b_n z^n$ are two
formal power series, $A \preceq B$ means that $\forall n \in \mathbb{N}, \,  a_n \leq b_n$.
Among other results, the authors of \cite{AR05} established the following:
\begin{lemma} \label{LEM_EGF}
Let $H_{\ell}(z)$ be the exponential generating function (EGF for short) of $b$-uniform
connected hypergraphs with excess $\ell$. Define by $T(z)$ the EGF
of labeled rooted hypertrees. Then,
\beq
H_{-1}(z) = T(z) - \frac{(b-1) \, T(z)^{b}}{b!} \quad \mbox{ with } \quad
T(z) =  z \, \exp{\left( \frac{T(z)^{(b-1)}} {(b-1)!}\right)} =%
 z \frac{\partial H_{-1}(z)}{\partial z}\, .
\eeq
For any $\ell \geq 1$, $H_{\ell}$ satisfies
\beq  \label{EQ:WRIGHT_INEQUALITIES} 
  \frac{\lambda_{\ell}(b-1)^{2\ell}}{3\,\ell \, T(z)^{\ell} \, \theta(z)^{3\ell}}
-\frac{(\nu_{\ell}(b-2))(b-1)^{2\ell-1}}
{(3\,\ell-1)\, T(z)^{\ell} \, \theta(z)^{3\ell-1}} \preceq %
  H_{\ell}(z) \preceq %
\frac{\lambda_{\ell}(b-1)^{2\ell}} {3\,\ell \,T(z)^{\ell} \, \theta(z)^{3\ell}} \, ,
\eeq
where  $\lambda_{\ell} = 3 \, \left( \frac{3}{2} \right)^{\ell} \frac{\ell !}{2 \pi} 
\left( 1 + O\left( \frac{1}{\ell} \right)\right)$
 and $\nu_{\ell} = O(\ell \lambda_{\ell})$.
Furthermore, $\lambda_{\ell}$ is defined recursively by
$\lambda_0 = \frac{1}{2}$ and
\ben \label{EQ:LAMBDA}
\lambda_{\ell} & = & \frac 12 \lambda_ {\ell -1 } (3\ell -1) \
  +\frac12 \sum_{t=0}^{\ell-1} \lambda_t \lambda_{\ell-1-t}  \, ,
\qquad (\ell \geq 1) \,  .
\een
\end{lemma}

We also need the following result which 
has been proved independently by Karo\'nski and {\L}uczak in \cite{KL97} and 
Andriamampianiana and Ravelomanana in \cite{AR05}:
\begin{lemma}
For  $\ell \equiv \ell(n)$ such that $\ell = o\left( \sqrt[3]{\frac{n}{b}} \right)$
as $n \ten \infty$, the number of connected $b$-uniform hypergraphs built with $n$ vertices
and having excess $\ell$ satisfies
\ben \label{BIG_EQUATION}
& & \sqrt{ \frac{3}{2 \, \pi}} \quad %
\frac{ \Big(b-1\Big)^{\frac{\ell}{2}} \quad %
e^{\frac{\ell}{2}} \quad n^{n+\frac{3 \, \ell}{2} - {1 \over 2}}}
{ 12^{\frac{\ell}{2}} \, \ell^{\frac{\ell}{2}} \, 
\Big( (b-2)! \Big)^{\frac{n+\ell}{b-1}} }
\, \exp{\left( \frac{n}{b-1} -n \right)} %
 \left(1 + O\left( \frac{1}{\sqrt{\ell}}\right) %
+ O\left(\sqrt{\frac{b \, \ell^3}{n}}\right) \right)  \, . \cr
& & \, 
\een
\end{lemma}
Observe that setting $b=2$ in (\ref{BIG_EQUATION}), we retrieve the asymptotical results of
Sir E. M. Wright for connected graphs in his fundamental paper \cite{Wr80}.

\section{Hypertrees and unicyclic components}
As typical examples, let us work with unicyclic components.
We will compute the expected number of
transitions $-1 \ten 0$. That is the number of times unicyclic connected components
(i.e. $0$-components) are created. We will also investigate the number of times
unicyclic components merge with hypertrees growing in size
but staying with the same complexity (excess $0$). In these
directions, we have the following result~:

\begin{theorem} \label{THM_CREAT_UNICYCLIC}
As $n \ten \infty$, on the average a $b$-uniform random hypergraph has
 about $\frac{1}{3} \log{n}$ dynamic unicyclic components. 
The expected number of static $0$-components is $\sim {\frac {\sqrt {2}{\pi }^{3/2} {24^{1/6}}}
{6 \, \Gamma  \left( \frac{5}{6} \right) } \, {(b-1)^{1/3}} \, n^{1/3} }
\approx 1.975 \, (b-1)^{1/3} \, n^{1/3}$.
\end{theorem}

\begin{proof} The creation of unicyclic components can be
obtained only by adding an edge joining
 $2$ distinct vertices inside the same hypertree with
$(b-2)$ other vertices from $(b-2)$ distinct hypertrees 
(to complete the edge). 

The number of such constructions is therefore given by
the coefficients of the following EGF~:
\beq
{C'}_0(z) = \frac{\Big(\Vz H_{-1}(z) \Big)^{(b-2)}}{(b-2)!} %
\times \left(\frac{\Vz^2-\Vz}{2} \Big(H_{-1}(z)\Big)\right) \, ,
\eeq
where the combinatorial
operator $\Vz = z \frac{\partial}{\partial z}$ corresponds to
 marking a vertice of the hypergraph in order to distinguish it from the
others. For instance, we refer the reader to 
Bergeron, Labelle and Leroux  \cite{BLL98} for the use of distinguishing/marking and 
pointing in combinatorial species.
Recall that the EGFs are as described briefly in Lemma \ref{LEM_EGF} (see also the
Appendix),  using $\Vz H_{-1}(z) = T(z)$ and $\Vz T(z) = \frac{T(z)}{1- T(z)^{(b-1)}/(b-2)!}$
we find
\beq
{C'}_0(z) = \frac{T(z)^{b-2}}{2\, (b-2)!} %
\left( \frac{T(z)}{1- \frac{T(z)^{(b-1)}}{(b-2)!}} - T(z) \right)
 = \frac{1}{2} \left( \frac{1}{1- \frac{T(z)^{(b-1)}}{(b-2)!}} - T(z) - 1\right) \, .
\eeq
We also have (such expansions are similar to those in \cite{KP89})
\beq
 \frac{1}{1- \frac{T(z)^{(b-1)}}{(b-2)!}} = %
\sum _{k=0}^{\infty }{\frac {{k}^{k}}{k!\,
 \coeff{  \left( b-2 \right) ! } ^{k}}} \, {z}^{ \left( b-1 \right) k} \, .
\eeq
Denoting by $\rho'((b-1)k,k)$ the number of manners to label a unicyclic 
component with $(b-1)k$ vertices and with a distinguished edge such that 
its deletion will leave a forest of hypertrees, we thus have
\beq
\rho'\Big((b-1)k,k\Big) = ((b-1)k)! \coeff{z^{(b-1)k}} \, {C'}_0(z) %
\sim \frac{ ((b-1)k)! \, {k}^{k} }{2 \, k!\, \coeff{  \left( b-2 \right) ! } ^{k}}
\eeq
(where if $A(z) = \sum_n a_n z^n$ then $\coeff{z^n}A(z) = a_n$).

Next, using Lemma \ref{LEMME_BINOMIAL_INTEGRAL} with the above equation, after
nice cancellations and summing other all possible values of $k$, we get
\ben
& & \sum_{k=1}^{\frac{n}{(b-1)}} \rho'\Big((b-1)k,k\Big) 
 {n \choose (b-1)k} \int_{0}^{1} %
t^{k-1} (1-t)^{ {n \choose b} - {n-(b-1)k \choose b}  - k} \, dt \cr
& & \sim \sum_{k=1}^{\frac{n}{(b-1)}} \frac{1}{2 k} 
\times \exp{\left( - \frac{(b-1)^4 \, k^3}{24 \, n^2} \right)} 
 \sim
\frac{1}{2} \int_{1}^{n/(b-1)} \frac{1}{x} \, e^{- (b-1)^4 \, x^3/24 n^2} \, dx \, .
\een
To estimate the last integral, we write
\ben
& & \int_{1}^{n/(b-1)} \frac{1}{x} \, e^{- (b-1)^4 \, x^3/24 n^2} \, dx \, =
\int_{1}^{n^{2/3}/(b-1)^{4/3}} \frac{1}{x}%
\left(1+O\left(\frac{(b-1)^4 x^2}{n^2}\right)\right) dx \cr
& + & O\left( \int_{n^{2/3}/(b-1)^{4/3}}^{n/(b-1)} %
\frac{1}{x} \, e^{- (b-1)^4 \, x^3/24 n^2} \, dx \right) \sim
\log{n^{2/3}} + O(1) \, .
\een
Thus, the expected number of creations of unicyclic components is $\sim \frac{1}{3}\, \log{n}$.
which completes the proof of the first part of the theorem. To prove the second part, we have
to investigate the number of static $0$-components, that is the number of times
$0$-components merge with hypertrees by the transition $0 \ten 0$.
The EGF of unicyclic components with a distinguished edge
such that its suppression will leave a unicyclic component and 
a set of $(b-2)$ rooted hypertrees is given by
\beq
{C''}_0(z) = \frac{T(z)^{b-2}}{(b-2)!} \, \Vz \Big(H_{0}(z)\Big) =
\frac{T(z)^{b-2}}{(b-2)!} %
\left( \frac{(b-1) \,  T(z)^{b-1}}{2 \, (b-2)! \, \theta^2} %
- \frac{ (b-1) \, T(z)^{b-1}}{ 2\, (b-2)! \, \theta} \right) \, 
\eeq
where $\theta = 1 - T(z)^{b-1}/(b-2)!$.
Denote by $\rho''((b-1)k,k)$ the number of manners to label a unicyclic 
component with $(b-1)k$ vertices and with a distinguished edge such that 
its deletion will leave a $0$-component with a forest of rooted
hypertrees, we claim that
\beq
\rho''\Big((b-1)k,k\Big) = ((b-1)k)! \coeff{z^{(b-1)k}} \, {C''}_0(z) %
\sim \sqrt{\frac{ \pi \, (b-1)^3}{8}} \frac{k^{k(b-1) + 1/2}}{e^{k(b-2)}} \, 
\left( \frac{ (b-1)^{k(b-1)}}{ \coeff{(b-2)!}^{k}} \right) \, .
\eeq
(We omit the details, since the full proof involves singularity analysis \cite{FS+} of the
EGF ${C''}_0$ described above.)

Now, using Lemma \ref{LEMME_BINOMIAL_INTEGRAL} and summing over $k$
after some cancellations, the computed expectation is about
\ben
& & \sum_{k=1}^{n/(b-1)} \sqrt{\frac{\pi}{8}} (b-1) \, \frac{1}{k^{1/2}} \, 
 e^{- (b-1)^4 \, k^3/24 n^2} \sim \sqrt{\frac{\pi}{8}} (b-1)
\int_{1}^{n/(b-1)}  \frac{e^{- (b-1)^4 \, x^3/24 n^2} \, dx}{\sqrt{x}} \cr
& & \sim 1/6\,{\frac {\sqrt {2}{\pi }^{3/2} {24^{1/6}}}
{\Gamma  \left( 5/6 \right) } \, {(b-1)^{1/3}} \, n^{1/3} } %
\approx 1.974748319\cdots \, (b-1)^{1/3} \, n^{1/3} \, .
\een
\end{proof}

Note here that the result stated in Theorem \ref{THM_CREAT_UNICYCLIC}
(humbly) generalizes the ones of Janson in \cite{Ja2000} since by setting
$b=2$, we retrieve his results concerning unicyclic (graph) components.

Next, we can investigate the number of vertices that ever belong to
$0$-components. According to the above computations,
the expected number of vertices added to $\mathcal{V}_0$ 
 for the creation of such unicyclic components  (transition $-1 \ten 0$)
is about
\beq
\frac{1}{2} \sum_{k=1}^{n/(b-1)} %
(b-1) \,  e^{- (b-1)^4 \, k^3/24 n^2} \sim \frac{1}{6} \, %
\frac{24^{1/3} \, \Gamma(1/3) }{(b-1)^{1/3}}  \, n^{2/3} \, .
\eeq
Whenever the excess $\ell$ is fixed, that is $\ell = O(1)$, the methods
developped here for unicyclic components can be generalized, using
analytical tools such those in \cite{FS+}.
Thus, we now turn on components with higher complexities.

\section{Multicyclic components with extremal complexities}
In this section, we focus on the creation and growth of components
of higher complexity. First, we will compute the expectations of
the number of creations of $\ell$-components for $\ell \geq 1$.
To this purpose, we need several intermediate lemmas.

Define $h_{n}(\a,\beta)$ as follows
\beq \label{DEF_HNLY}
\frac{1}{{T(z)}^{\a} \, \left(1-\frac{T(z)^{b-1}}{(b-2)!} \right)^{3\a+\beta}}
= \sum_{n\geq 0}h_{n}(\a ,\beta) \frac{z^n}{n!} \, .
\eeq
The following lemma is an application of the saddle point method 
\cite{De Bruijn,FS+} which is well suited to cope with 
our analysis~:
\def\b{\beta} 
\begin{lemma} \label{LEMMA_SADDLE}
Let $\a\equiv \a(n)$ be such that $\a(b-1) \ten 0$ but ${\a(b-1)n \over \ln{n}^2} \ten \infty$ and
let $\beta$ be a fixed number. Then $h_{n}(\a n,\beta)$ defined in (\ref{DEF_HNLY})
satisfies
\ben \label{FORMULA_SADDLE}
 h_{n}(\a n,\beta) 
 & =&  \frac{n!}{\sqrt{2 \pi n \Big(b-1\Big)} \,  \Big( (b-1)! \Big)^{\frac{\a n+n}{b-1}} } %
\, \Big( 1-(b-1)u_0 \Big)^{(1-\b)} \cr 
&\times & \exp{\left(n\Phi(u_0)\right)} %
\left( 1+O\Big(\sqrt{\a(b-1)}\Big)+O\Big(\frac{1}{\sqrt{\a(b-1)n}}\Big) \right) \, ,
\een
where 
\ben \label{PHI_U0}
\Phi(u) &=&  u - \left(\frac{\a+1}{b-1}\right) \ln{u} - 3 \, \a \ln{(1-(b-1)u)} \cr
 u_0 &=& {\frac {3/2\,\a b-\a+1-1/2\,\sqrt\Delta}{b-1}} \qquad \mbox{with }%
\, \Delta =  {9\,{\a}^{2}{b}^{2}-12\,{\a}^{2}b+12\,\a b+4\,{\a}^{2}-12\,\a}.
\een 
\end{lemma}

\begin{proof}  One can start with Cauchy's integral formula.
Note that the radius of convergence of the series $T(z)$ is given by
$\sqrt[(b-1)]{ (b-2)! } \, \exp{\left( - 1/(b-1) \right)}$. 
We make the substitution $u=T(z)^{(b-1)}/(b-1)!$ and
get successively 
\ben
T(z)  & = & \sqrt[(b-1)]{(b-1)! \, u} \, , \quad %
z  =  \sqrt[(b-1)]{(b-1)! \, u} \, e^{-u} \quad {\mbox{ and }} \cr
dz & = & \left(\frac{1}{(b-1) \, u } - 1 \right) \, %
\Big( (b-1)! \, u \Big)^{\frac{1}{(b-1)} } 
 \,  e^{-u} \, du \, .
\een
From the Cauchy integral formula, we then obtain
\beq
h_n(\a\,n,\b) = \frac{n!}{2 \pi i \Big( (b-1)!\Big)^{(\a \, n+n)/(b-1)} } %
\oint \frac{ \left(1- (b-1) u\right)^{1-\b} }{(b-1) \, u} \, %
\exp{\left(n \Phi(u)\right)} \, du \, ,
\label{COL_DEUX}
\eeq
where $\Phi(u) = u - \left(\frac{\a+1}{b-1}\right) \ln{u} - 3 \, \a \ln{(1-(b-1)u)}$.
The big power in the integrand, \textit{viz.} $\exp{(n \Phi(u))}$, suggests us to use the
 saddle point method. Investigating the roots of $\Phi'(u) = 0$, we find
two saddle points,
$u_0 = {\frac {3/2\,\a b-\a+1-1/2\,\sqrt\Delta}{b-1}}$  and 
$u_1  =  {\frac {3/2\,\a b-\a+1+1/2\,\sqrt\Delta}{b-1}}$
 with $\Delta =  {9\,{\a}^{2}{b}^{2}-12\,{\a}^{2}b+12\,\a\, b+4\,{\a}^{2}-12\,\a}$.
Moreover, we have
$\Phi''(u)={\frac {\a+1}{ \left( b-1 \right) {u}^{2}}}+3\,{\frac {\a \left( -b+1 
 \right) ^{2}}{ \left( 1- \left( b-1 \right) u \right) ^{2}}}$
so that for $u \protect\notin \{0, 1/(b-1)\}$, $\Phi''(u) >0$.
The main point of the application of the saddle
point method here is that $\Phi^{'}(u_0)=0$ and
$\Phi^{''}(u_0) > 0$, hence $n \Phi(u_0 \exp{(i\tau)})$
is well approximated by $n\Phi(u_0) - n {u_0}^2 \Phi^{''}(u_0) \frac{\tau^2}{2}$
in the vicinity of $\tau = 0$.
If we integrate (\ref{COL_DEUX}) around a circle 
passing vertically through $u=u_0$ in the $z$-plane, we obtain
\beq
h_{n}(\a n, \b) = \frac{n!}{2 \pi \, \Big( (b-1)!\Big)^{(\a n+n)/(b-1)} } %
\int_{-\pi}^{\pi} \frac{\left(1- (b-1) u_0 e^{i \tau} \right)^{1-\b}}%
{(b-1)} \, %
\exp{\left(n \Phi(u_0 e^{i \tau}) \right)} \, d\tau \, 
\label{COL_QUATRE}
\eeq
where
\begin{equation}
\Phi(u_0 e^{i\tau}) = u_0 \cos \tau + i u_0 \sin \tau %
- \frac{\a+1}{b-1}\ln u_0 - i\frac{\a+1}{b-1}\tau - 3 \a \ln (1-(b-1) u_0 e^{i \tau}) \, \, .
\end{equation}
Denoting by $\EuFrak{Re}(z)$ the real part of $z$, if 
$f(\tau)=\EuFrak{Re}( \Phi(u_0 e^{i \tau}))$ we have
\beq
f(\tau) 
 = u_0 \cos \tau - \frac{\a+1}{b-1}\ln{u_0} - 3 \a \ln{u_0} - 3 \a\ln{(b-1)} %
 - \frac{3 \a}{2} \ln{\Big(1 + \frac{1}{(b-1)^2 u_0^2} - \frac{2 \cos \tau}{(b-1)u_0} \Big)} \, .
\eeq
It comes
\beq
f'(\tau) = \frac{d}{d \tau}  \EuFrak{Re}(h(u_0 e^{i\tau})) = %
- u_0 \sin \tau - %
\frac{ 3 \a \sin \tau}{u_0(b-1) + \frac{1}{(b-1)u_0} - 2 \cos \tau} \, .
\eeq
Therefore,  if $\tau=0$ $f^{'}(\tau) = 0$. Also, $f(\tau)$ is
a symmetric function of $\tau$ and in 
$\left[ -\pi, -\tau_0 \right] \cup \left[ \tau_0, \pi \right]$, for
any given $\tau_0 \in (0, \, \pi)$, and $f(\tau)$ takes its maximum value for
$\tau = \tau_0$. Since $|\exp(\Phi(u))| = \exp( \EuFrak{Re}(\Phi(u)) )$,
when splitting
the integral in (\ref{COL_QUATRE}) into three parts, viz.
``$\int_{-\pi}^{-\tau_0} + \int_{-\tau_0}^{\tau_0}
 + \int_{\tau_0}^{\pi}$'', we know that it suffices to integrate
from $-\tau_0$ to $\tau_0$, for a convenient value of
$\tau_0$, because
the others can be bounded by the magnitude of the integrand
at $\tau_0$. 
In fact, we have
$\Phi(u_0 e^{i \theta}) = \Phi(u_0) + \sum_{p\geq 2} \phi_p (e^{i\theta} -1)^p$
with  $\phi_p = \frac{{u_0}^p}{p!} \Phi^{(p)}(u_0)$.
We easily compute $ \Phi^{(p)}(u_0) = (-1)^p (p-1)! \Big( \frac{\a+1}{(b-1){u_0}^{p}} +
\frac{3 \a (1-b)^p}{{(1-(b-1)u_0)}^{p}}\Big)$, for $p \geq 2$. 
Whenever $\a b \ten 0$, we have
$(b-1)u_0 =1-\sqrt {3\,(b-1) \, \a}+ \left( 3/2\,b-1 \right) \a+O \left( {b}^{3/2
}{\a}^{3/2} \right)$.
Therefore, we obtain after a bit of algebra
\beq
| \phi_p | \leq O \left( \frac{2^p}{ {\a}^{\frac{p}{2}-1}  (b-1)^{\frac{p}{2}}}\right) \,,  %
 \qquad \mbox{as } \a(b-1) \rightarrow 0 \, .
\eeq
On the other hand,
\begin{equation}
| e^{i\tau} - 1 | = \sqrt{ 2(1- \cos \tau)} < \tau %
\,, \qquad \tau > 0 \, .
\end{equation}
Thus, the summation  can be bounded for values of
$\tau$ and $\a$ such that $\tau \rightarrow 0$, 
$\a b \ten 0$ ($\a \ten 0$) but $\frac{\tau}{\sqrt{\a}} \rightarrow 0$ and we have
\beq
\Big\arrowvert \sum_{p \geq 4} \phi_p (e^{i\tau} - 1)^p \Big\arrowvert %
 \leq \sum_{p \geq 4} | \phi_p \tau^p |
 \leq  \sum_{p \geq 4} %
O\Big( \frac{2^p \tau^p}{\a^{\frac{p}{2}-1} (b-1)^{\frac{p}{2}}}\Big) %
       =  O\Big( \frac{\tau^4}{\a(b-1)} \Big) \, .
\eeq
It follows that for $\tau \rightarrow 0$, 
$\a(b-1) \rightarrow 0$ and $\frac{\tau}{\sqrt{\a(b-1)}} \rightarrow 0$,
$\Phi(u_0 e^{i\tau})$ can be rewritten as
\ben
\Phi(u_0 e^{i\tau}) & = & \Phi(u_0) - \frac{1}{(b-1)}%
\left( 1 - \frac{\sqrt{\a}}{\sqrt{3(b-1)}} \, \frac{3b-4}{2} +%
\frac{(9b^2-12b+4)}{12(b-1)} \a  \right) \tau^2 \cr
&-& %
\frac{i}{(b-1)} \left( 1 - \frac{(3b-4) \sqrt{\a}}{2\, \sqrt{3(b-1)}} %
 +  \frac{(9b^2-12b+4)}{12(b-1)} \, \a \right) \tau^3 + %
O\left( \frac{\tau^4}{\a(b-1)} \right) \, .
\een
Therefore, if $\a(b-1) \rightarrow 0$ but 
$\frac{\a(b-1)n}{{(\ln{n})}^2} \rightarrow \infty$, if we let
$\tau_0 = \frac{\ln n}{\sqrt{n \, u_0^2 \Phi''(u_0)}}$ 
(with $u_0^2 \Phi''(u_0) = \frac{2}{b-1} + O(\sqrt{\a(b-1)})$) 
we can remark (as already said) that  it suffices to 
integrate (\ref{COL_QUATRE}) from $- \tau_0$ to
$\tau_0$, using the magnitude of the integrand at $\tau_0$
to bound the resulting error. 
\noindent The rest of the proof is 
now standard application of the saddle point method
(see for instance De Bruijn \cite[Chapters 5 \& 6]{De Bruijn})
and is omitted in this extended abstract. After a bit of algebra,
one gets the formula (\ref{FORMULA_SADDLE}).
\end{proof}

\begin{lemma} \label{SKETCHED1}
Let $a=k(b-1) - \ell$. Denote by $c_{\ell}(a,k)$ the number of
manners to label an $\ell$-component with $a$ vertices 
such that one edge  -- whose deletion will suppress the occurrence 
of the created $\ell$-component --  is distinguished among the others.
As $\ell$ tends to $\infty$ with the number of vertices $a$ 
such that  $\ell = o\left( \sqrt[3]{\frac{a}{b}} \right)$ then
\beq
c_{\ell}(a,k) =   a! \coeff{z^a} %
\left( \frac{1}{2} \, \frac{(3\ell)\, (b-1)^{2\ell} \, \lambda_{\ell-1} }%
{T(z)^{\ell}  \theta^{3\ell+1}} \right) \, 
\times
\left( 1 + O\left( \frac{1}{\sqrt{\ell}}  %
+ O\left(\sqrt{\frac{b \, \ell^3}{a}}\right) %
\right) \right) \, ,
\eeq
where $\theta = 1- T(z)^{b-1}/(b-2)!$ and the sequence
$(\lambda_{\ell})$ is defined with (\ref{EQ:LAMBDA}).
\end{lemma}

\noindent
\textit{Sketch of proof.} The proof given in this extended abstract is sketched.
The main ideas are as follows. The inequalities given by
equation (\ref{EQ:WRIGHT_INEQUALITIES}) in Lemma \ref{LEM_EGF} tell us that when $\ell$ is large,
the main constructions that lead to the creation of a new $\ell$-component arises 
a.a.s. from picking two distinct vertices in an $(\ell-1)$-component
and joining them by an edge with $(b-2)$ set of rooted hypertrees. 
Such constructions are counted by
\[
\left( \frac{\Vz^2 - \Vz}{2} H_{\ell-1}(z) \right) \times \frac{T(z)^{b-2}}{(b-2)!} \, .
\]
Using again (\ref{EQ:WRIGHT_INEQUALITIES}) with (\ref{FORMULA_SADDLE}), one can show
 that the coefficient of the latter EGF has the same asymptotical behaviour as the following one
\[
\frac{3\ell \, (b-1)^{2\ell} \lambda_{\ell-1}}{2\, T(z)^{\ell} \theta^{3\ell+1}} \, .
\]
(The error terms being the same as those given by the saddle-point Lemma 
\ref{LEMMA_SADDLE} above.)
\ENDPROOF

We then have the following result~:
\begin{theorem}
As $\ell, b \ten \infty$ with $n$ but such that $\ell = o\left( \sqrt[3]{\frac{n}{b}} \right)$,
 the expected number of creations of $\ell$-component is $\sim$ 1 and
the expected number of vertices that ever belong to an $\ell$-component
is about $\left( 12 \ell (b-1)\right)^{1/3} n^{2/3}$. 
Thus, $\qE\coeff{\VMAX} = O\left( \ell^{1/3} (b-1)^{1/3} n^{2/3}\right)$.
\end{theorem}

\noindent
\textit{Sketch of the proof.}
The proof of this Theorem is a combination of Lemmas \ref{LEMMA_SADDLE},\ref{SKETCHED1}
  and \ref{LEMME_BINOMIAL_INTEGRAL}
 together with the asymptotic value of $\lambda_{\ell}$ given in Lemma \ref{LEM_EGF}
and with the fact that 
\[
\sum_{k=0}^{n/(b-1)} k^u \exp{\left( - \frac{(b-1)^4 k^3}{24 \, n^2} \right)} \sim
\frac{1}{3} \, \frac{24^{\frac{u+1}{3}} \, n^{\frac{2(u+1)}{3}} }{(b-1)^{\frac{4(u+1)}{3}} } \,
\Gamma\left( \frac{u+1}{3} \right) \, .
\]

\bibliographystyle{plain}

\end{document}